\documentclass[10pt,twocolumn]{article}

\usepackage{latex8}
\usepackage{times}

\usepackage{amsmath}
\usepackage{thanks}

\usepackage{psfig}

\usepackage{url}
\usepackage{amssymb}

\usepackage{subfigure}
\usepackage{graphicx}

\usepackage{color}

\usepackage{graphics}
\usepackage{latexsym}

\newcommand{\oldbfe}[1]{\begin{bfseries}\emph{#1}\end{bfseries}}

\newcommand{\fa}{\mbox{$\forall$}}

\newcommand{\NI}{\noindent}

\newcommand{\II}{\vspace{2 mm}}

\newcommand{\szkew}[1]{\relax \setbox0=\hbox{\kern -24pt $\displaystyle#1$\kern 0pt }%
\box0}
{\catcode`\@=11 \global\let\ifjusthvtest@=\iffalse}

\newcounter{oldmycaption}

\input{epsf}

\def\smallromani{\renewcommand{\theenumi}{\roman{enumi}}
\renewcommand{\labelenumi}{(\theenumi)}}

\begin{document}

\date{}

\title{A Distributed Platform for Mechanism Design}

 \author{Krzysztof R. Apt \thanksref{t1}\thanksref{t2}\\
 \and Farhad Arbab  \thanksref{t1}\thanksref{t3} \\
 \and Huiye Ma \thanksref{t1}
 }

 \thankstext{t1}{CWI, Amsterdam,  Kruislaan 413, 1098 SJ Amsterdam, The Netherlands}
 \thankstext{t2}{University of Amsterdam}
 \thankstext{t3}{University of Leiden, The Netherlands}

\maketitle

\thispagestyle{empty}

\begin{abstract}
  We describe a structured system for distributed mechanism design. It
  consists of a sequence of layers.  The lower layers deal with the
  operations relevant for distributed computing only, while the upper
  layers are concerned only with communication among players,
  including broadcasting and multicasting, and distributed decision
  making.  This yields a highly flexible distributed system whose specific
  applications are realized as instances of its top layer. 
  

This design supports
  fault-tolerance, prevents manipulations and makes it possible to
  implement distributed policing. The system is implemented in Java.
  We illustrate it by discussing a number of implemented
  examples.
\end{abstract}

\section{Introduction}
\label{sec:intro}

\subsection{Background and motivation}

Mechanism design is an important area of economics.  It aims at
realizing economic interactions in which
desired social decisions result when each agent is interested in
maximizing his utility.
%
The traditional approach relies on the existence of a
central authority,
who collects the information from the players,
computes the decision and informs the players about the outcome and
their taxes.

Recently, in a series of papers distributed mechanism design was
suggested as a realistic alternative for the applications based on the
Internet. In this setting no central authority exists and the decisions
are taken by the players themselves.
The challenge here is to appropriately combine the techniques of
distributed computing with those that deal with the matters specific
to mechanism design, notably rationality 
and truth-telling. 

\subsection{Related work}

A number of recent papers deal with different aspects of distributed
mechanism design. 
%
An influential paper
\cite{FS02} introduced the notion of distributed algorithmic
mechanism design emphasizing the issues of computational complexity
and incentive compatibility in distributed computing.
Next, \cite{PS04} studied the distributed
implementations of the VCG mechanism. However, in their approach there
is still a center that is ultimately responsible for selecting and
enforcing the outcome.

The authors of \cite{Shneidman2004S} considered the problem of creating distributed system
specifications that will be faithfully implemented in networks with
rational (i.e., self-interested) nodes so that no node will choose to
deviate from the specification. 
%
Researchers of \cite{Petcu2006M} introduced the first distributed
implementation of the VCG mechanism.  The only central authority
required was a bank that is in charge of the computation of taxes.
The authors also discussed a method to
redistribute some of the VCG payments back to players.

\subsection{Contributions}

In this paper we propose a platform for distributed mechanism design.
Our work is closest to \cite{Petcu2006M} whose approach is based on
distributed constraint programming.  In contrast, our approach builds
upon a very general view of distributed programming, an area that
developed a variety of techniques appropriate for the problem at hand.

Our platform is built out of a number of layers.  This leads to a
flexible, hierarchical design in which the lower layers are concerned
only with the matters relevant for distributed computing, such as
communication and synchronization issues, and 
the upper layers that deal with the relevant aspects of the
mechanism design, such as computation of the desired decision and taxes.
Any specific application is realized simply
as an \emph{instantiation of a top layer}.  
This layered architecture offers a number of novel features and
improvements to the approach of \cite{Petcu2006M}, to wit

\begin{itemize}
\item we deal with a larger class of mechanisms, notably Groves
  mechanisms.  Additionally, we can easily tailor our platform to
  other tax-based mechanisms, such as Walker mechanism (see
  \cite{Wal81}),
\vspace{-3mm}

\item we support open systems in which the number of players can be unknown,
  \vspace{-3mm}

\item the bank process of \cite{Petcu2006M} is replaced by a weaker
  \oldbfe{tax collector} process. It is needed only for the mechanisms
  that are not balanced, wherein it is a passive process used only to receive
messages to collect the resulting deficit,
\vspace{-3mm}

\item \oldbfe{fault-tolerance} is supported, both on the message transmission
  level and on the player processes level,
\vspace{-3mm}

\item a multi-level protection against \oldbfe{manipulations} is provided,
\vspace{-3mm}

\item our platform makes it possible to implement \oldbfe{distributed policing}
  that provides an alternative to a `central enforcer' whose
  responsibility is to implement the outcome decided by the agents and
  collect the taxes (see, e.g., \cite[ page 366]{FSS07}).
\vspace{-3mm}
  
\end{itemize}

Fault-tolerance at the
mechanism design level means that the final decision and taxes
can be computed even after some of the processes that broadcast the
player's types crash: the other processes then still can proceed.
This is achieved by the duplication of the computation by all players.
Such a redundancy is common in all approaches to fault-tolerance (and
also used to prevent manipulations, see \cite[ page 366]{FSS07}). It
was intentionally avoided in \cite{Petcu2006M} which aimed at
minimizing the overall communication and computation costs. In our
approach it allows the fastest process to `dominate' the computation
and move it forward more quickly.

Our platform can be easily customized to real-life applications. Using
it players can engage in joint decision making by dynamically forming
a network with no central authority, in which they know neither their
neighbours nor the size of the network.  Also it can be used for a
repeated distributed decision making process, each round involving a
different group of interested players.  This design is implemented in
Java.

Finally, a few words about the paper organization.  In the next
section we review the basic facts about the tax-based mechanisms,
notably the Groves family of mechanisms. 
Then in Section \ref{sec:our}
we discuss the issues that need to be taken care of when moving from
the centralized tax-based mechanisms to distributed ones and what
approach we took to tackle these issues. The details of our design and
implementation are provided in Section \ref{sec:details}.

Next, in Section \ref{sec:manip}, we discuss three important
advantages of our design: security, distributed policing and
fault-tolerance.  In Section \ref{sec:examples} we discuss
a number of examples of mechanisms that we implemented using our
system. They include Vickrey auction with redistribution, two types of
auctions and a sequential mechanism design. Then, in Section
\ref{sec:conc} we provide conclusions.

\section{Preliminaries: mechanism design}
\label{sec:classical}

We recall here briefly tax-based mechanisms,
see, e.g., \cite[ Chapter 23]{MWG95}.
Assume a set of \textbf{\textit{decisions}} $D$, a set $\{1, \ldots, n\}$ of players, 
for each player a set of \textbf{\textit{types}} $\Theta_i$ and a \textbf{\textit{utility function}}
$
v_i  : D \times \Theta_i \rightarrow \cal{R}.
$
In this context a type is some private information known only to the player, for example
a vector of player's valuations of the items for sale in a multi-unit auction.

A \textbf{\textit{decision rule}} is a function $f: \Theta \rightarrow D$, where
$\Theta := \Theta_1 \times \cdots \times \Theta_n$.  We call the tuple
\[
(D, \Theta_1, \ldots, \Theta_n, v_1, \ldots, v_n, f)
\]
a \textbf{\textit{decision problem}}.

A decision rule $f$ is called \textbf{\textit{efficient}} if for all $\theta \in
\Theta$
\[
f(\theta)\in {\rm argmax}_{d \in D} \sum_{i=1}^n v_i(d,\theta_i), 
\]
and \textbf{\textit{strategy-proof}} (or \textbf{\textit{incentive compatible}})
if for all $\theta \in \Theta$, $i \in \{1,\ldots,n\}$ and
$\theta'_i \in \Theta_i$
\[
v_i(f(\theta_i, \theta_{-i}), \theta_i) \geq
v_i(f(\theta'_i, \theta_{-i}), \theta_i),
\]
where $\theta_{-i} := (\theta_1, \ldots, \theta_{i-1}, \theta_{i+1}, \ldots, \theta_n)$ and
$(\theta'_i, \theta_{-i}) := (\theta_1, \ldots, \theta_{i-1}, \theta'_i, \theta_{i+1},$ $\ldots, \theta_n)$.

In mechanism design one is interested in the ways of inducing the
players to announce their true types, i.e., in transforming the
decision rules to the ones that are strategy-proof.  In
\oldbfe{tax-based} mechanisms this is achieved by extending the
original decision rule by means of \textbf{\textit{taxes}} that are
computed by the central authority from the vector of the received
types, using players' utility functions.

Given a decision problem, in the classical setting, 
one considers then the following sequence of events, where $f$ is a
given, publicly known, decision rule:
\begin{enumerate} \smallromani

\item each player $i$ receives a type $\theta_i$,
\vspace{-3mm}

\item each player $i$ announces to \emph{the central authority} a type $\theta'_i$;
this yields a joint type $\theta' := (\theta'_1, \ldots, \theta'_n)$,
\vspace{-3mm}

\item the central authority then computes the decision $d := f(\theta')$
and the tax vector $t := g(\theta')$, where 
$g : \Theta \rightarrow {\cal R}^n$ is
a given function, and communicates to each player $i$ the decision $d$ and his tax $t_i$.
\vspace{-3mm}

\item the resulting utility for player $i$ is then
$u_i(d,t) := v_i(d, \theta_i) + t_i$.
\end{enumerate}

Each \oldbfe{Groves mechanism} is obtained using $g(\theta') :=
(t_1(\theta'), \ldots, t_n(\theta')), $ where for all $i \in \{1,
\ldots, n\}$

\begin{itemize}

\item 
$
t_i(\theta') :=
  h_i(\theta'_{-i}) + \sum_{\substack{j=1\\j\not=i}}^{n} v_j(f(\theta'), \theta'_j),
$
\item
$
h_i: \Theta_{-i} \rightarrow \mathbb{R}
$
is an arbitrary function.

\end{itemize}

The importance of the Groves mechanisms is revealed by the following crucial result.
\II

\NI
\textbf{Groves Theorem}
Suppose the decision rule $f$ is efficient. Then in each Groves mechanism 
the expanded decision rule 
$
(f, g): \Theta \rightarrow D \times {\cal R}^n
$
is strategy-proof w.r.t.~
the utility functions $u_1, \ldots, u_n$.
\II




When for a given tax-based mechanism for all $\theta'$ we have $\sum_{i = 1}^{n} t_i(\theta') \leq 0$
(respectively, $\sum_{i = 1}^{n} t_i(\theta') = 0$),
the mechanism is called \textbf{\textit{feasible}} (respectively, \textbf{\textit{budget balanced}}).
A special case of a feasible Groves mechanism, called \textbf{\textit{Vickrey-Clarke-Groves mechanism}}
(in short \textbf{\textit{VCG}}) is obtained by
using 
$
h_i(\theta'_{-i}) := - \max_{d \in D} \sum_{j \neq i} v_j(d, \theta'_j).
$
So then
\[
t_i(\theta')  := \sum_{j \neq i} v_j(f(\theta'), \theta'_j) - \max_{d \in D} \sum_{j \neq i} v_j(d, \theta'_j).
\]


\section{Our approach}
\label{sec:our}

In our approach we relax a number of the assumptions made when
introducing mechanism design.  More specifically we assume that

\begin{itemize}
\item there is no central authority, \vspace{-3mm}
  
\item players interested in participating in a specific mechanism
  register to join an open system wherein that mechanism runs,
  \vspace{-3mm}

\item the players whose registration is accepted (request the
  underlying distributed communication layer to) broadcast their type
  to other players in the system,
\vspace{-3mm}

\item once a registered player learns that he has received the types
   from all registered players, he computes the decision and the taxes,
   sends this information to other registered players and terminates
   his computation.
\vspace{-3mm}

\end{itemize}

As in all cited works, we also assume that there is no collusion among the players.
This leads to an implementation of the mechanism design by means of
distributed processes.
The
computation of the decision and of the taxes is carried out by the
players themselves.






In our approach each player is
represented by a process, in short a \oldbfe{player process}.  A
player who wishes to join a specific mechanism (e.g., an auction) must
register with a \oldbfe{local registry}.  
Local registries are linked together in a network that satisfies the
full reachability condition.

Once player process registration is successful, it joins the network
of (registry and player) processes wherein a generic
\oldbfe{broadcast} command is available.  The implementation of this
command relies only on the assumption that for each pair of players
there is a path of neighbouring processes connecting them.  This
allows us to deal with arbitrary network topologies in a simple way.
The broadcast messages are transmitted through paths managed in a lower layer
which the player processes \emph{cannot access}. This automatically prevents
manipulation by player processes of messages originating
from or destined for other players, a problem
pointed out e.g. in \cite[ page 366]{FSS07}.

Each player process after broadcasting the player's type
participates in a \textbf{\textit{distributed termination detection
    algorithm}} (see, e.g., \cite{MC98})
the aim of which is to learn whether all players have
indeed broadcast their types.  
If this algorithm detects termination, the player process
knows that he indeed received all types, and in particular can determine at
this stage the number of players.
More generally, we use the distributed termination detection algorithm to detect
the end of each \emph{phase} of the distributed computation: registration, type broadcast, etc.,
i.e., for \oldbfe{barrier synchronization}, see, e.g., \cite{And05}.

Each
player process uses the same, publicly known, decision rule $f$ that
he learns, for example from a public bulletin board, and as a result computes
the same decision.
%
Further, each player process applies $f$ to the same input $\theta'$
and computes \emph{the same} \textbf{\textit{tax scheme}}
$tax(t_1), \ldots tax(t_n)$ from
the tax vector $(t_1, \ldots, t_n)$, where $tax(t_j)$ specifies the
amounts that player $j$ has to pay to other players and possibly the tax
collector
 from his tax $t_j$.  All tax schemes $tax(t_1), \ldots
tax(t_n)$ then determine `who pays how much to whom'.  
The tax collector process is only needed for the mechanisms
that are not budget balanced.



\section{Implementation}
\label{sec:details}

\begin{figure}[htbp]
\centerline{\psfig{figure=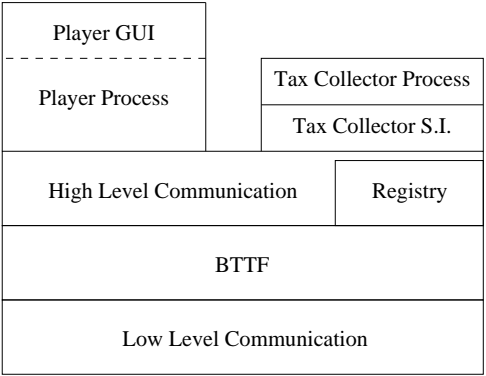,height=5cm}}
\caption{\label{F.Layers} Implementation architecture}
\end{figure}

Our distributed mechanism design system is implemented in Java and consists of
about 12.5 K lines of Java code. The
implementation follows the guidelines explained in the previous
section.  Figure~\ref{F.Layers} shows the overall architecture of our
system and the different layers of software used in its
implementation. 
Each entity in this
architecture communicates, either through function calls or method
invocations, \emph{only} with its adjacent entities.  Specific
applications are realized by instantiating the crucial player process
layer.

\vspace{-3mm}

\paragraph{Low Level Communication}
\label{subsec:LLC}

The Low Level Communication (LLC) layer supports (1) locally
generated, globally unique process identifiers, and (2) reliable
non-order-preserving, asynchronous, targeted communication, exclusively
through the exchange of passive messages between processes.  The only
means of communication between processes in LLC is through message
passing, where no transfer of control takes place when messages are
exchanged.

%
Successful send simply means that
the message has been dispatched on its way to its specified target.


\vspace{-3mm}

\paragraph{BTTF}
\label{subsec:BTTF}

The Back To The Future (BTTF) layer implements a message
efficient, fault-tolerant distributed termination detection (DTD)
algorithm, on top of the LLC layer. The details of the BTTF DTD
algorithm 
lie beyond the scope of this paper and will be described elsewhere.
%

All DTD algorithms determine termination using waves of
special control messages, called tokens.  They also require the
designation of a single process as the \oldbfe{initiator}, which is
responsible for initiating the waves.
%
In the BTTF algorithm, the initiator is anonymous, i.e., no process
(other than the initiator) knows who the initiator is.  

The DTD functionality provided by the BTTF layer can be used for
barrier synchronization as well as for termination detection.

\vspace{-3mm}

\paragraph{High Level Communication and Registry}

The High Level Communication (HLC) layer provides indirect,
anonymous communication among the players in a distributed
system.  It includes a number of local registries whose mutual
connectivity supports the full connectivity of the players
necessary for broadcast.  A player must sign-in at a local
registry, after which it can use the other operations provided by
the HLC layer to take part in the mechanism.







Each local registry is responsible for processing the registrations of
the player processes according to the assumed registration criteria.

\vspace{-3mm}

\paragraph{Player Process}
\label{subsec:player} 

Specific applications are implemented using this top layer. 
It is built on top of the HLC layer and is used to implement
specific actions of the players, in particular the computation
of the decisions and taxes. 

\vspace{-3mm}

\paragraph{Tax Collector Software Interface}

This layer is built on top of the HLC layer and provides the
counterparts of the functions of the HLC layer to deal with the tax
collector process registration.

\vspace{-3mm}

\paragraph{Tax Collector Process}

This layer is built on top of the Tax Collector Software Interface
layer and is used to implement the actions of the tax
collector which is in charge of collecting players' taxes.
We omit the details.







\vspace{-3mm}

\paragraph{Player GUI}

The interaction between the player (user) and the system is realized
in this interface.  The interaction is limited to the registration,
type submission and tax reception.

\section{Security, Distributed Policing and Fault-tolerance}
\label{sec:manip}

\begin{figure}[htbp]
\centerline{\psfig{figure=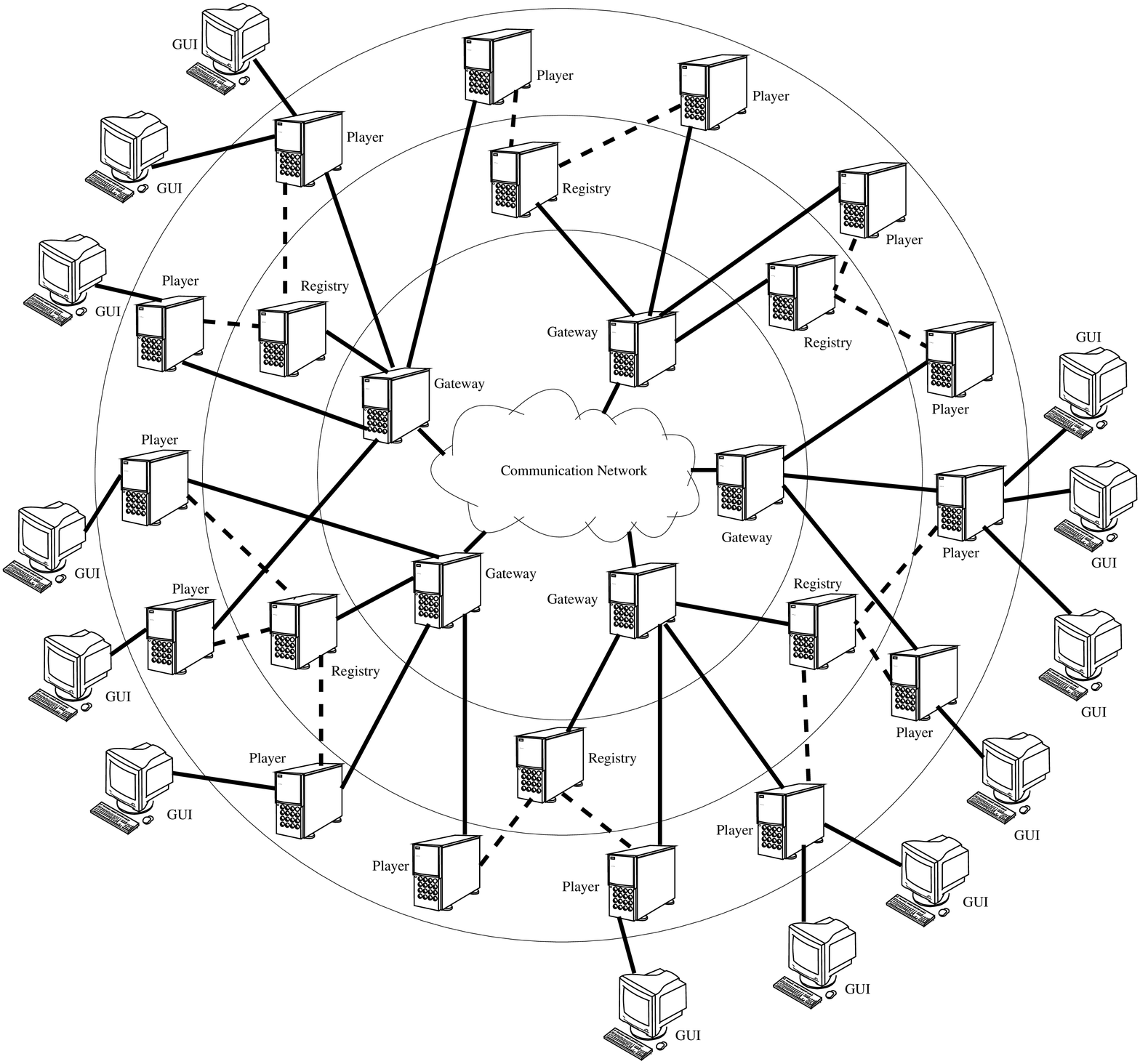,height=7cm}}
\caption{\label{P} A realization of the platform}
\end{figure}

Figure \ref{P} shows a mapping of the architectural elements described
in the previous section to `logical hosts'. The mapping is \oldbfe{dynamic} and
hence unknown to players.  In any concrete implementation one or more
such logical hosts can represent the same actual physical host. The
\oldbfe{communications network}, represented by the cloud shape,
interconnects a number of hosts to provide the functionality described
in the LLC layer in Section \ref{sec:details}.  The specific hosts
connected to this network that concern us are a set of \oldbfe{gateway
  hosts that run the BTTF and the HLC layers}.

The ring of hosts around the core in Figure \ref{P} contains the set of
\oldbfe{hosts that run the local registries}.  Every local registry 
has a primary connection to a gateway host in the core.  Thus, the full
reachability of the gateway hosts in the core ensures full reachability
among local registries.

The next ring of hosts in Figure \ref{P} contains \oldbfe{hosts that run
player processes}.  Each player process establishes an initial link
(dotted lines) with a local registry (whose address it obtains from a
local forum) to register.  As part of this registration process, its
local registry provides the address of a gateway host with which the
player process then establishes its primary communication link (solid
lines) for the rest of the game.  
Finally, the outermost ring in Figure \ref{P} consists merely of
\oldbfe{computers that run GUI programs} that link to their respective
player processes.

This `ring structure' provides multiple forms of protection against
manipulations by the players.  
The only
messages that pass through a player process are the ones originating
from or destined for that specific player.  Furthermore, the end users have physical access only to the outermost
hosts that run the GUI programs, which severely restricts the range of
their potentially dangerous actions.  Finally, the separation of the
GUI programs from the player processes allows us to run the latter on
hosts to which end users do not have physical access.
Users can trust the security of the
messages they exchange through a `public' communications system, by
relying on the encryption of the messages using the public key
cryptography.

A dishonest user may attempt to alter the
code of a player process so that it sends to some players a falsified
decision or a falsified tax scheme.  By \oldbfe{policing} we mean here
a sequence of actions that will lead to the exclusion of such
processes (that we call \emph{dishonest}).  The qualification
`distributed' refers to the fact that the policing is done by the
player processes themselves, without intervention of any central
authority.  We call a player process \emph{honest} if it
multicasts a true tax scheme.

The difficulty in implementing distributed policing lies in the fact
that dishonest processes may behave inconsistently. To resolve this
problem we make use of registries that are assumed to be reliable.  We
then modify the sequence of actions of each player process so that it
always computes the decision and the tax scheme but sends them only to
its local registry.  The local registry then dispatches the tax scheme
on behalf of its sender to all player processes mentioned in the tax
scheme.  As a trusted intermediary, the registry ensures that the same
tax scheme is sent to all player processes involved, and that no
player process can send more than one tax scheme in a single phase.

The BTTF algorithm in the
BTTF layer detects persistent process failures.
The duplication of the computation by all
players allows us to easily modify the design to support
fault-tolerance at
the mechanism design level.

\section{Examples}
\label{sec:examples} 

We used our distributed mechanism design system in a number of test
cases that we now briefly describe.  Each of them, 
is implemented as an instantiation of the player process layer
described in Section \ref{sec:details}.
\vspace{-3mm}

\paragraph{Vickrey auction with redistribution}
\label{subsec:vickrey}

In Vickrey auction there is a single object for sale which is
allocated to the highest bidder who pays the second highest bid. We
implemented the proposal of \cite{Cav06} in which the highest bidder
redistributes some amounts from his payment to other players. This
minimizes the overall tax. Due to space limitations we omit the details.

\vspace{-3mm}

\paragraph{Unit demand auction}

In this auction multiple items are offered for sale. We
assume that there are $n$ players and $m \leq n$ items and that each player
submits a valuation for each item.  The items should be allocated in
such a way that each player receives at most one of them 
and the aggregated valuation is maximal.
This auction can be modelled as the following decision problem:


\begin{itemize}

\item  $D = \{f \mid f: {\cal P}(\{1, \ldots, m\}) \rightarrow \{1, \ldots, n\}, f  \mbox{ is 1-1}\}$,

\vspace{-3mm}

\item $\Theta_i = {\cal R}^{m}_{+}$; $(\theta_{i, 1}, \ldots, \theta_{i, m}) \in \Theta_i$ is a vector of
player $i$'s valuations of the items for sale,
\vspace{-3mm}

 \item 
$
v_i(d,\theta_i) := \left\{ \begin{array}{ll}
\theta_{i, j} & \textrm{if $d(j) = i$} \\
0 & \textrm{if $\neg \exists j \: d(j) = i$}
\end{array} \right.
$
\vspace{-3mm}

\item 
$
f(\theta') \in {\rm argmax}_{d \in D} \sum_{j \in dom(d)} \theta'_{d(j), j}.
$
\vspace{-3mm}



\end{itemize}

Decision rule $f$ is efficient, so Groves Theorem applies.
%
The decisions are computed using the Kuhn-Munkres
algorithm to compute the maximum weighted matching, where the weight
associated with the edge $(j,i)$ is the valuation for item $j$
reported by player $i$.  In our implementation we used the Java source
code available at \url{http://adn.cn/blog/article.asp?id=49}.
To compute tax for player $i$ according to the VCG mechanism
this algorithm needs to be used again, to 
compute the maximum weighted matching with
player $i$ excluded.

\vspace{-3mm}

\paragraph{Single minded auction}
\label{subsec:single}

In this auction studied in \cite{LCS02} there are $n$ players and $m$
items, with each player only interested in a specific set of items.
For simplicity we assume
that each player $i$ is only
interested in a subsequence $a_i, \ldots, b_i$ of the items
$1, \ldots, m$.
We model this as the following decision problem:

\begin{itemize}


\item  $D = \{f \mid f: {\cal P}(\{1, \ldots, m\}) \rightarrow \{1, \ldots, n\}\}$, 
\vspace{-3mm}

\item $\Theta_i = {\cal R}_{+}$; $\theta_i \in \Theta_i$ is player
  $i$'s valuation for the sequence $a_i, \ldots, b_i$ of the items,
\vspace{-3mm}

\item 
$
v_i(d,\theta_i) := \left\{ \begin{array}{ll}
\theta_{i} & \textrm{if $\fa j \in [a_i, \ldots, b_i] \: d(j) = i$} \\
0 & \textrm{otherwise}
\end{array} \right.
$
\vspace{-3mm}

\item 
$
f(\theta') \in {\rm argmax}_{d \in D} \sum_{i: d([a_i, \ldots, b_i]) = \{i\}} \theta'_{i},
$

where $d([a_i, \ldots, b_i]) = \{d(j) \mid j \in [a_i, \ldots, b_i]\}$.
\vspace{-3mm}

\end{itemize}

So, given an allocation $f \in D$ the goods in the set $\{k \mid f(k) = j\}$
are allocated to player $j$.
Decision rule $f$ is efficient and consequently Groves Theorem applies.
The computations of the decision and of the
taxes 
involve constructions of the maximum weighted matchings that are
computed using a dynamic programming algorithm, details of which are omitted.

\vspace{-3mm}

\paragraph{Other applications}

To test the versatility of our approach
we also implemented a number of other examples,
including
%
Vickrey auction,
decision making concerned with public projects (see 
 \cite[ Chapter 23]{MWG95}), sequential Groves mechanisms
studied in \cite{AE07}, 
and Walker mechanism of \cite{Wal81}.
%
%
%
%
The latter mechanism is not an instance of Groves mechanism and
implements the decision not in dominant strategies but
in a Nash equilibrium (see, e.g., \cite{MWG95}).  

\section{Conclusions}
\label{sec:conc}


We described here the design
and implementation of a platform that supports distributed mechanism
design.
%
We believe that the proposed platform clarifies how the design of
systems supporting distributed decision making can profit from sound
principles of software engineering, such as separation of concerns and
hierarchical design.  

We found that the division of the software into layers resulted in a
flexible design that could be easily customized to specific mechanisms.
For example, our distributed implementation of Vickrey
auction required modification of a module of only 60 lines of code.
Additionally, this layered architecture offers a multi-level
protection against manipulations, distributed policing and supports
fault-tolerance.

We also provided evidence that software engineering in the area of
multi-agent systems can profit from the techniques developed in the
area of distributed computing, for example broadcasting in an
environment with an unknown number of processes, distributed
termination and barrier synchronization.


\paragraph{Acknowledgments}

 We thank Kees Blom and Han Noot for providing us with the software
 used in the system, and Vangelis Markakis for proposing the
 appropriate algorithms for the unit demand and single minded auctions.
 The work of Huiye Ma was funded by the NWO project DIACoDeM, No
 642.066.604.

\bibliography{/ufs/apt/bib/e,/ufs/apt/bib/sin02}
\bibliographystyle{latex8}

\end{document}